\documentclass[12pt]{article}
\usepackage{graphicx}
\usepackage{amsmath}
\usepackage{amssymb}
\usepackage{amstext}
\usepackage{amscd}
\usepackage{amsfonts}
\usepackage{appendix}
\DeclareMathAlphabet{\EuFrak}{U}{euf}{m}{n}
\DeclareMathAlphabet{\EuScript}{U}{eus}{m}{n}

\title{{\bf  Features of the fractional diffusion-advection equation}}
\author{M.C.Rocca$^{1,2}$,A.R.Plastino$^{3}$,\\A.Plastino$^{1,2}$, G. L. Ferri$^4$, A. de Paoli$^{1,2}$\\\\
\small{$^1$Departamento de F\'{\i}sica, Fac. de Ciencias Exactas},\\
\small{Universidad Nacional de La Plata}\\
\small{C.C. 67 (1900) La Plata, Argentina}\\
\small{$^2$IFLP-CCT-CONICET-C.C 727 (1900) La Plata. Argentina}\\
\small{$^3$ CeBio y Secretaria de Investigacion,}\\
\small{Univ. Nac. del Noroeste de la Prov. de Bs. As.,}\\
\small{ UNNOBA and CONICET, R. Saenz Pe\~{n}a 456, Junin,
Argentina}\\ $^4$ \small{Faculty of Exact and Natural Sciences, La Pampa National University} \\
\small{Uruguay 151, Santa Rosa, La Pampa, Argentina}}

\date{\today}

\begin{document}

\maketitle

\begin{abstract}

We advance an exact, explicit form for the solutions to the
fractional diffusion-advection equation. Numerical analysis of
this equation shows that its solutions resemble power-laws.


\end{abstract}

\newpage

\renewcommand{\theequation}{\arabic{section}.\arabic{equation}}

\section{Introduction}

Advection constitutes a transport mechanism of either a substance
or a conserved property by a fluid. This process originates in the
fluid's bulk motion. For instance,  the transport of silt in a
river by bulk water flowing downstream. Thermodynamic  advected
quantities are energy and enthalpy.  Any substance or conserved,
extensive quantity can experience advection by a fluid that
contains it. Advection  does not include transport  by molecular
diffusion.    In this work we analyze some features of an
evolution equation characterizing the combined effects of i)
advection and ii) a diffusion process  described by fractional
partial derivatives. The ensuing fractional advection-diffusion
equation has been recently applied to the study of the transport
of energetic particles in the interplanetary environment
\cite{SS2011,TZ2011,ZP2013,C95,LE2014}.

There is a considerable body of evidence, from data
collected by spacecrafts like {\it Ulysses} and {\it Voyager 2},
indicating that the transport of energetic particles in the
turbulent heliospheric medium is superdiffusive
\cite{PZ2007,PZ2009}. Considerable effort has been devoted in
recent years to the development of superdiffusive models for the
transport of electrons and protons in the heliosphere
\cite{SS2011,TZ2011,ZP2013}. This kind of transport regime
exhibits a power-law growth of the mean square displacement of the
diffusing particles, $\langle \Delta x^2 \rangle \propto
t^{\alpha}$, with $\alpha > 1$ (see, for instance, \cite{SZ97}).
The special case $\alpha = 2$ is called ballistic transport. The
limit case $\alpha \to 1$ corresponds to normal diffusion,
described by the well-known Gaussian propagator. The energetic
particles detected by the aforementioned probes are usually
associated with violent solar events like solar flares. These
particles diffuse in the solar wind, which is a turbulent
environment than can be assumed statistically homogeneous at large
enough distances from the sun \cite{PZ2007}. This implies that the
propagator $P(x,x',t,t')$, describing the probability of finding
at the space time location $(x,t)$ a particle that has been
injected at $(x',t')$, depends solely on the differences $x-x'$
and $t-t'$. In the superdiffusive regime the propagator
$P(x,x',t,t')$ is not Gaussian, and exhibits power-law tails. It
arises as solution a non local diffusive process governed by an
integral equation that can be recast under the guise of a
diffusion equation where the well-known Laplacian term is replaced
by a term involving fractional derivatives \cite{C95}. Diffusion
equations with fractional derivatives have attracted considerable
attention recently (see \cite{LTRLGS2014,
LSSEL2009,RLEL2007,stern,LMASL2005} and references therein) and have
lots of potential applications \cite{MK2000,P2013}. In particular,
the observed distributions of solar cosmic ray particles are often
consistent with power-law tails, suggesting that a superdiffusive
process is at work.

A proper understanding of the transport of energetic
particles in space is a vital ingredient for the analysis of
various important phenomena, such as the propagation of particles
from the Sun to our planet or, more generally, the acceleration
and transport of cosmic rays. The superdiffusion of particles in
interplanetary turbulent environments is often modelled using
asymptotic expressions for the pertinent non-Gaussian propagator,
which have a limited range of validity. A first step towards a
more accurate analytical treatment of this problem was recently
provided by Litvinenko and Effenberger (LE) in \cite{LE2014}. LE
considered solutions of a fractional diffusion-advection equation
describing the diffusion of particles emitted at a shock front
that propagates at a constant upstream speed $V_{sh}$ in the solar
wind rest frame. The shock front is assumed to be planar, leading
to an effectively one-dimensional problem. Each physical quantity
depends only on the time $t$ and on the spatial coordinate $x$
measured along an axis perpendicular to the shock front.

In the present contribution we re-visit the fractional
diffusion-advection equation and provide {\sf explicit}, exact
closed analytical  solutions, without approximations. We also
undertake a numerical analysis that shows the these solutions
resemble power-laws.

\section{Formulation of the Problem}

\setcounter{equation}{0}

We deal with the equation
\begin{equation}
\label{ep2.1}
\frac {\partial f} {\partial t}=\kappa\frac
{{\partial}^{\alpha} f} {\partial |x|^{\alpha}}+a\frac {\partial
f} {\partial x}+\delta(x),
\end{equation}
where $t>0$ and $f(x,t)$ is the distribution function for solar
cosmic-rays transport. Here the fractional spatial derivative is
defined as \cite{LE2014}
\begin{equation}
\label{ep2.2}
\frac {{\partial}^{\alpha} f} {\partial
|x|^{\alpha}}= \frac {1} {\pi}\sin\left(\frac {\pi\alpha}
{2}\right)\Gamma(\alpha+1) \int\limits_0^{\infty}\frac
{f(x+\xi)-2f(x)+f(x-\xi)} {{\xi}^{\alpha+1}}\;d\xi.
\end{equation}
To solve this equation one appeals to the the Green function
governed by the equation:
\begin{equation}
\label{ep2.3}
\frac {\partial {\cal G}} {\partial t}=\kappa\frac
{{\partial}^{\alpha} {\cal G}} {\partial
|x|^{\alpha}}+\delta(x)\delta(t).
\end{equation}
With this Green function, the solution of (\ref{ep2.1}) can be
expressed as
\begin{equation}
\label{ep2.4}
f(x,t)=\int\limits_0^t {\cal G}(x+at^{'},
t^{'})\;dt^{'}.
\end{equation}
In this work we obtain the solutions of Eqs. (\ref{ep2.1}) and
(\ref{ep2.3})  {\bf using  distributions as main tools
 \cite{guelfand1}.}

For our task we use, as a first step, the solution obtained in
\cite{LE2014} for the Green function through the use of the Fourier
Transform given by
\begin{equation}
\label{ep2.5}
\hat{{\cal G}}(k,t)=\frac {1} {2\pi}
\int\limits_{-\infty}^{\infty}{\cal G}(x,t) e^{-ikx}\;dx,
\end{equation}
from which we obtain for $\hat{{\cal G}}$:
\begin{equation}
\label{ep2.6}
\hat{{\cal G}}(k,t)=-\kappa|k|^{\alpha}\hat{{\cal
G}}(k,t)+\frac {1} {2\pi}\delta(t),
\end{equation}
whose solution is
\begin{equation}
\label{ep2.7}
\hat{{\cal G}}(k,t)=\frac {H(t)} {2\pi} e^{-\kappa
|k|^{\alpha}t},
\end{equation}
where $H(t)$ is the Heaviside's step function.

\section{Explicit general solution of the equation}

\setcounter{equation}{0}

From (\ref{ep2.7}) we have for $\hat{{\cal G}}$
\begin{equation}
\label{ep3.1}
\hat{{\cal G}}(k,t)=\frac {H(t)} {2\pi} e^{-\kappa
|k|^{\alpha}t}= \frac {H(t)} {2\pi}
\sum\limits_{n=0}^{\infty}\frac {(-1)^n{\kappa}^nk^{\alpha n} t^n}
{n!},
\end{equation}
and, invoking  the inverse Fourier transform
\[{\cal G}(x,t)=\frac {H(t)} {2\pi}
\int\limits_{-\infty}^{\infty}
e^{-\kappa |k|^{\alpha}t}e^{ikx}\;dk=\]
\begin{equation}
\label{ep3.2}
\frac {H(t)} {2\pi} \sum\limits_{n=0}^{\infty}\frac
{(-1)^n{\kappa}^n t^n} {n!} \left[\int\limits_0^{\infty}k^{\alpha
n}e^{ikx}\;dx + \int\limits_0^{\infty}k^{\alpha
n}e^{-ikx}\;dx\right].
\end{equation}
Fortunately, we can find in the classical book of \cite{guelfand1}
 the results for the two integrals of (\ref{ep3.2}). We
obtain
\begin{equation}
\label{ep3.3}
{\cal G}(x,t)=\frac {H(t)} {2\pi}
\sum\limits_{n=0}^{\infty}\frac {(-1)^n{\kappa}^n t^n} {n!}
\Gamma(\alpha n+1) \left[\frac {e^{i\frac {\pi} {2}(\alpha n+1)}}
{(x+i0)^{\alpha n + 1}} + \frac {e^{-i\frac {\pi} {2}(\alpha
n+1)}} {(x-i0)^{\alpha n + 1}} \right].
\end{equation}
Using now (\ref{ep2.4}) we have for $f$
\[f(x,t)=\int\limits_0^t
{\cal G}(x+at^{'}, t^{'})\;dt^{'},\] so that one can write
\[f(x,t)=\frac {1} {2\pi}
\sum\limits_{n=0}^{\infty}\frac {(-1)^n{\kappa}^n} {n!}
\Gamma(\alpha n+1)\times\]
\begin{equation}
\label{ep3.4}
 \int\limits_0^t \left[\frac {e^{i\frac {\pi}
{2}(\alpha n+1)}} {(x+at^{'}+i0)^{\alpha n + 1}} + \frac
{e^{-i\frac {\pi} {2}(\alpha n+1)}} {(x+at^{'}-i0)^{\alpha n + 1}}
\right]t^{'n}\;dt^{'}.
\end{equation}
According to Eq. (\ref{a1}) of the  Appendix, where $t>0$, we now
obtain for $f$, invoking hypergeometric functions
$F(\alpha n+1,2;3;z)$ and Beta functions ${\cal B}(1,n+1)$,
\[f(x,t)=\frac {1} {2\pi}\sum\limits_{n=0}^{\infty}
 \frac {(-1)^n{\kappa}^nt^{n+1}} {n!}\Gamma(\alpha n+1)
{\cal B}(1,n+1)\times\]
\[\left[\frac {e^{i\frac {\pi} {2}(\alpha n + 1)}}
 {(x+i0)^{\alpha n + 1}}
F\left(\alpha n+1,n+1;n+2;-\frac {at} {x+i0}\right)+\right.\]
\begin{equation}
\label{gral} \left.\frac {e^{-i\frac {\pi} {2}(\alpha n + 1)}}
 {(x-i0)^{\alpha n + 1}}
F\left(\alpha n+1,n+1;n+2;-\frac {at} {x-i0}\right)\right].
\end{equation}
This is the general solution of Eq.(\ref{ep2.1}) for the initial
condition $f(x,0)=0$.

\begin{figure} [htb]
\centerline{{\scalebox{0.55} {\includegraphics{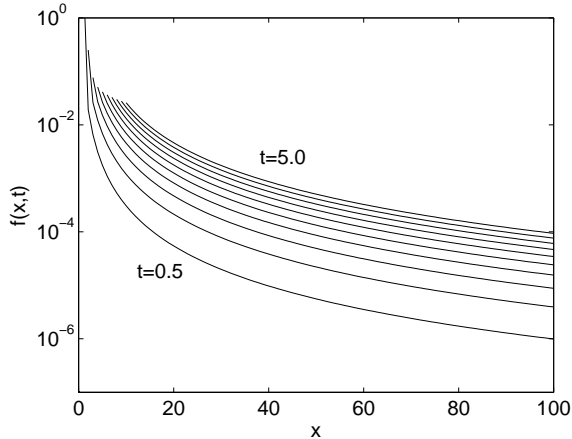}}}}
 \caption{Typical graph for solutions of Eq. (\ref{gral}),
 for different times $t$} ($x$ in arbitrary units). \label{Fig1}
\end{figure}


\newpage

\section{Results}

Fig. 1 depicts a typical graph for solutions of Eq. (\ref{gral}).
We do this for different times $t$.  The main result of this
contribution is that these curves are very well approximated by
power laws. Figs. 2 ($\kappa=0.01$), 3 ($\kappa=0.5$), and 4
($\kappa=0.2$) show how the above solutions can be adjusted by
power laws of the form $1/x^q$: the corresponding lines rotate,
for fixed $\alpha=3/2$, in the plane $f(x,t)$ vs. $1/x^q$ when
$at$ changes, as the plots clearly illustrate. In all our figures
we have taken $a=1$ and $\alpha=3/2$.

\noindent Figs. 5, 6, and 7 show how the dependence of $q$ with
$\alpha$ can, in turn, be adjusted in simple fashion. In Figs. 5
and 6, straight lines are employed. In Fig. 7 we attempt a spline
adjustment. It is evident that a weak diffusion-regime exists.
This regime was conjectured  in Ref. \cite{LE2014} by appeal to an
approximate theoretical treatment of the fractional
diffusion-advection equation and here amply confirmed by our exact
approach.

\noindent What about other $\alpha$ values? The situation does not
change. Power-law adjustment continues to be possible. Figs. 8 and
9 illustrate this issue.

\section{Conclusions}

 We have provided an explicit analytical solution for and
 advection-diffusion equation involving fractional derivatives in
 the diffusion term. This equation governs the diffusion of
 particles in the solar wind injected at the front of a shock
 that travels at a constant upstream speed $V_{sh}$ in the solar
 wind rest frame. The shock is assumed to have a planar front,
 leading to a problem with an effective one dimensional geometry,
 where all the relevant quantities depend solely on time and on
 the coordinate $x$ measured along an axis perpendicular to the
 front.

 We obtained the exact solution of the above mentioned
 equation in the $x$-configuration space (besides the
 associated formal solution in the $k$-space related
 to the previous one via a Fourier transform).

 We conclude from our analysis that the solutions of the fractional diffusion-advection
 equations are essentially power laws, and have numerically found a
 ``law" for the behavior of the associated power-exponent $q$ with
 varying $\kappa$ via spline interpolation.

\newpage

\renewcommand{\thesection}{\Alph{section}}

\renewcommand{\theequation}{\Alph{section}.\arabic{equation}}

\setcounter{section}{1}

\section*{Appendix:Some properties of Hypergeometric Function}

\setcounter{equation}{0}

Using data from \cite{gra2} we have
\[\int\limits_0^t\frac {t^{'n}}
{(x+at^{'}\pm i\epsilon)^{\alpha n +1}}\;dt^{'}= \frac {t^{n+1}}
{(x\pm i\epsilon)^{\alpha n+1}}B(1,n+1)\times\]
\begin{equation}
\label{a3} F\left(\alpha n +1, n+1,n+2;-\frac {at} {x\pm
i\epsilon}\right).
\end{equation}
Then, we have
\[\lim_{\epsilon\rightarrow 0^+}\int\limits_0^t\frac {t^{'n}}
{(x+at^{'}\pm i\epsilon)^{\alpha n +1}}\;dt^{'}=
\lim_{\epsilon\rightarrow 0^+} \frac {t^{n+1}} {(x\pm
i\epsilon)^{\alpha n+1}}B(1,n+1)\times\]
\begin{equation}
\label{a2} \lim_{\epsilon\rightarrow 0^+} F\left(\alpha n +1,
n+1,n+2;-\frac {at} {x\pm i\epsilon}\right).
\end{equation}
Thus we obtain finally
\[\int\limits_0^t\frac {t^{'n}} {(x+at^{'}\pm i0)^{\alpha n +1}}\;dt^{'}=
\frac {t^{n+1}} {(x\pm i0)^{\alpha n+1}}B(1,n+1)\times\]
\begin{equation}
\label{a1} F\left(\alpha n +1, n+1,n+2;-\frac {at} {x\pm
i0}\right).
\end{equation}
\newpage

\begin{figure} [htb]
\centerline{{\scalebox{0.55} {\includegraphics{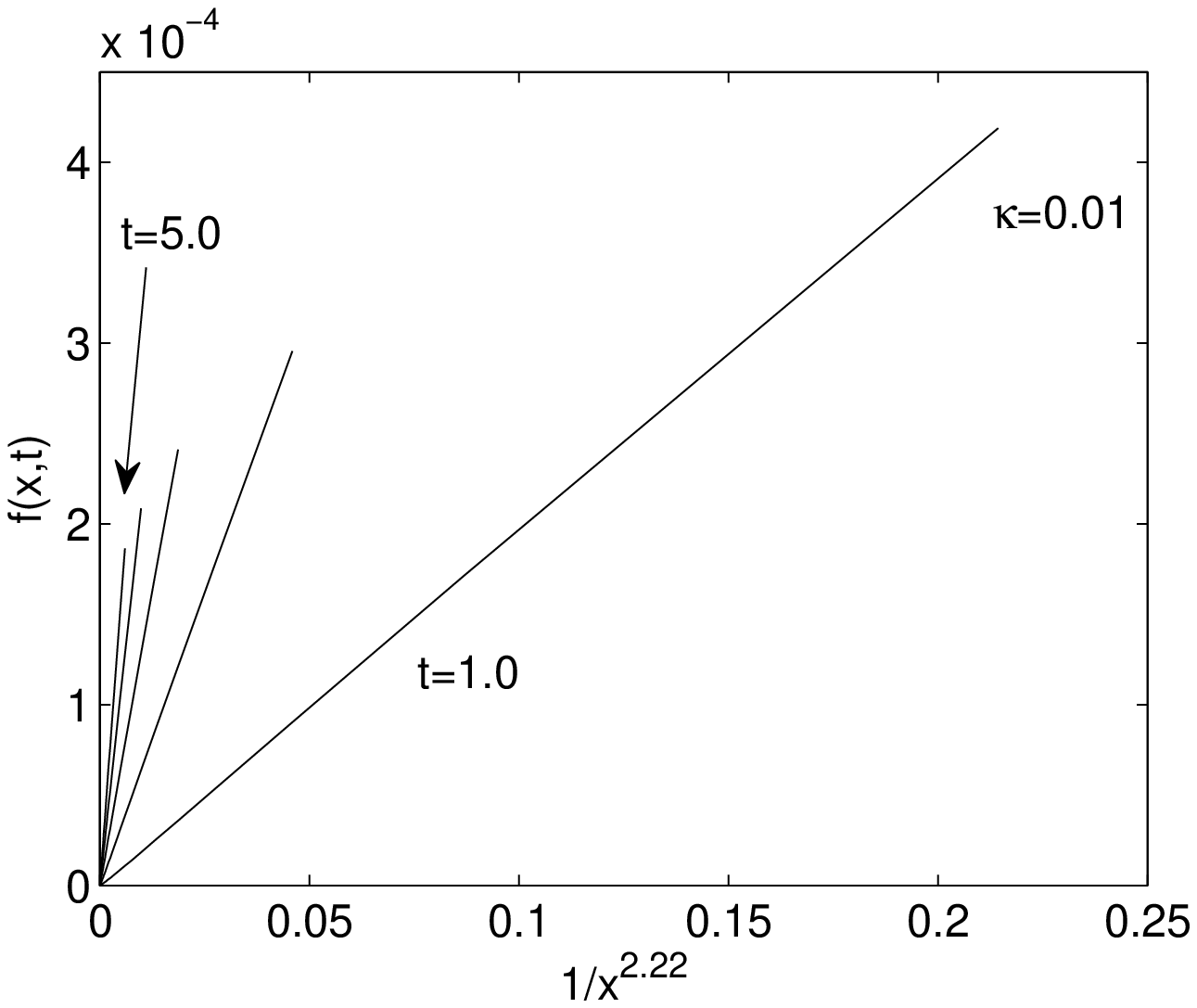}}}}
 \caption{Power-law fitting to the solutions of Eq. (\ref{gral}) for $\alpha=3/2$ $a=1$, and $\kappa=0.01$, at different times $t$ }
\label{Fig2}
\end{figure}

\newpage

\begin{figure} [htb]
\centerline{{\scalebox{0.55}
 {\includegraphics{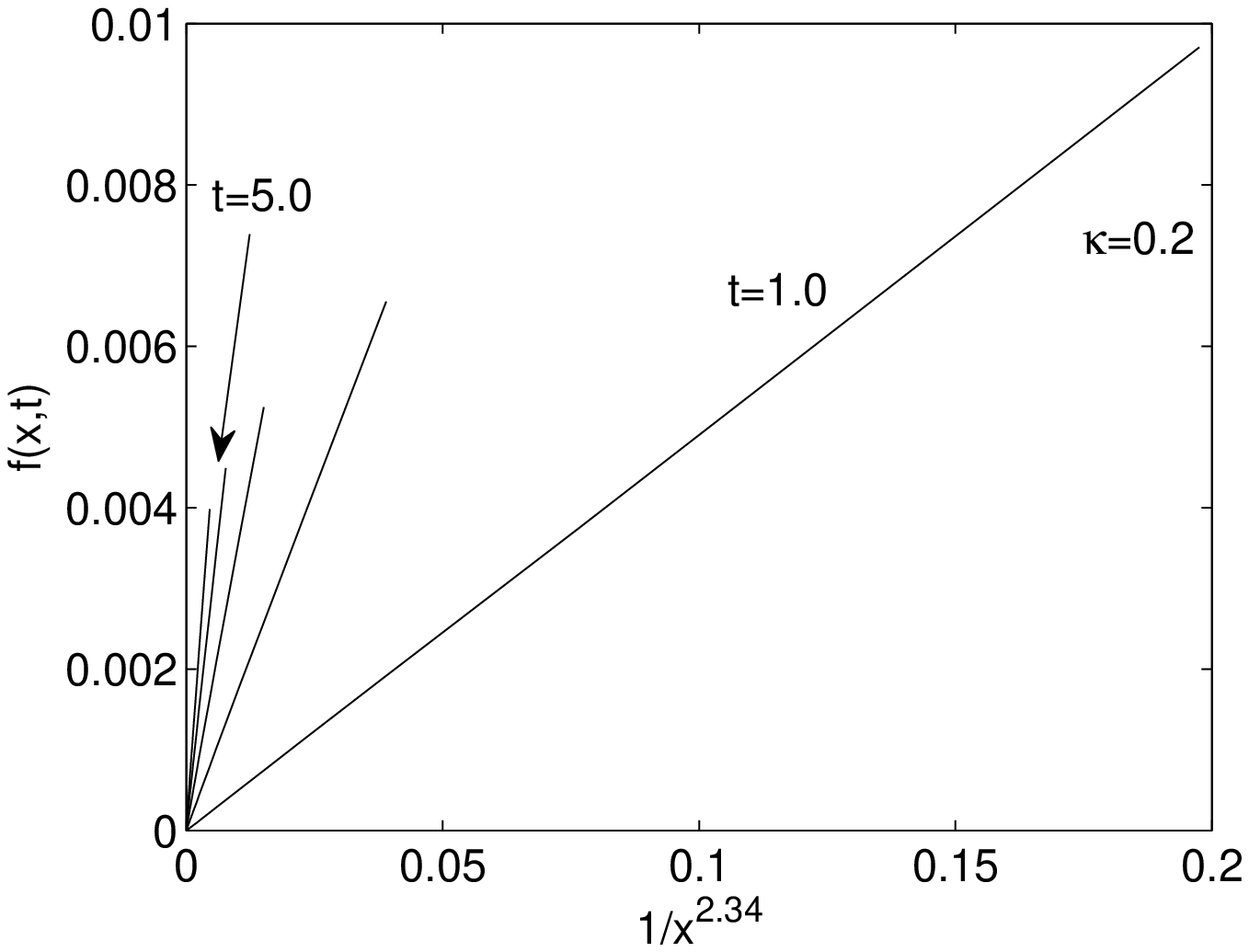}}}}
 \caption{Power-law fitting to the solutions of Eq. (\ref{gral}) for $\alpha=3/2$ $a=1$,  and $\kappa=0.5$, at different times $t$ }
\label{Fig3}
\end{figure}

\newpage

\begin{figure} [htb]
\centerline{{\scalebox{0.55} {\includegraphics{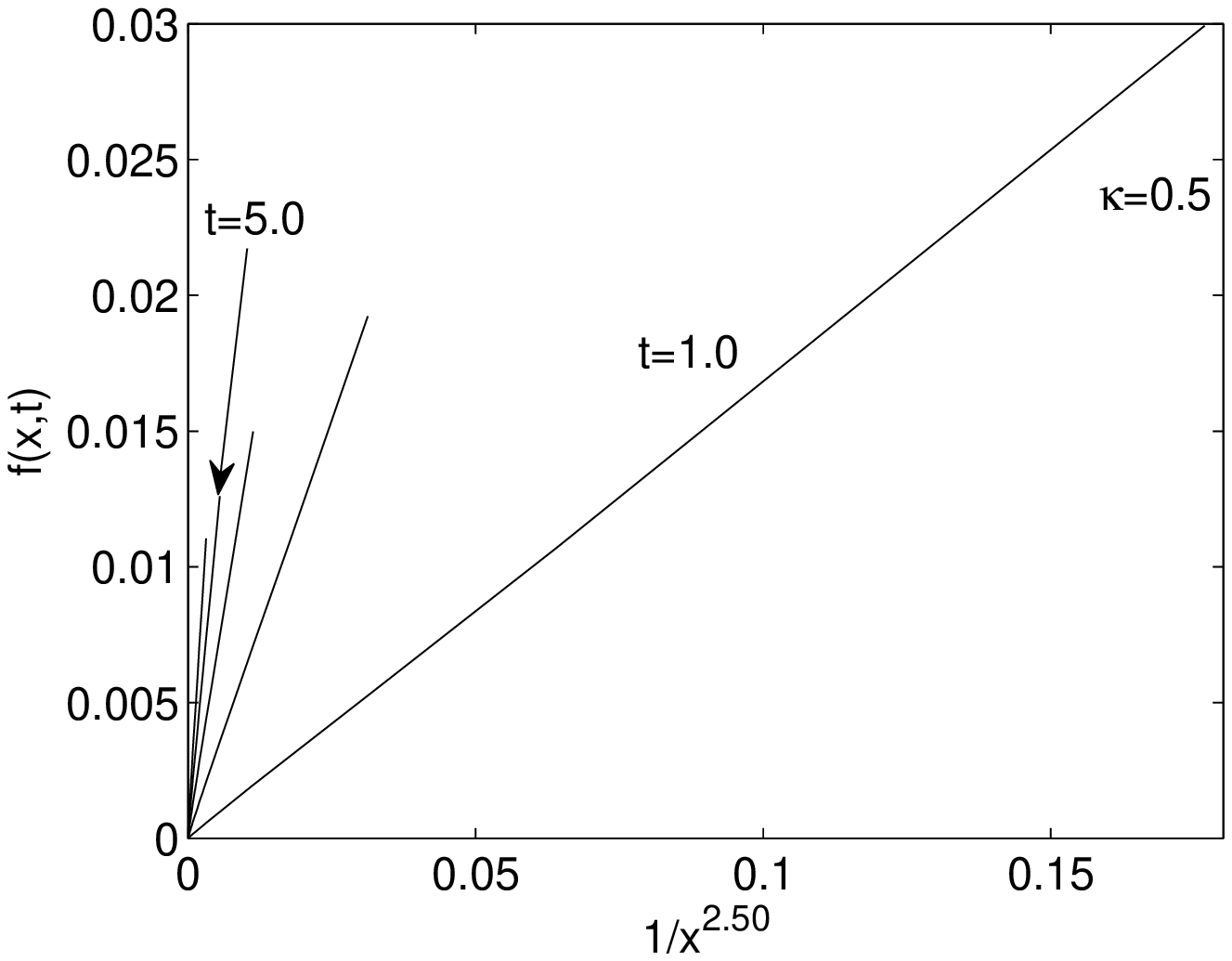}}}}
 \caption{Power-law fitting to the solutions of Eq. (\ref{gral}) for $\alpha=3/2$, $a=1$, and $\kappa=0.2$, at different times $t$ }
\label{Fig4}
\end{figure}

\newpage

\begin{figure} [htb]
\centerline{{\scalebox{0.55} {\includegraphics{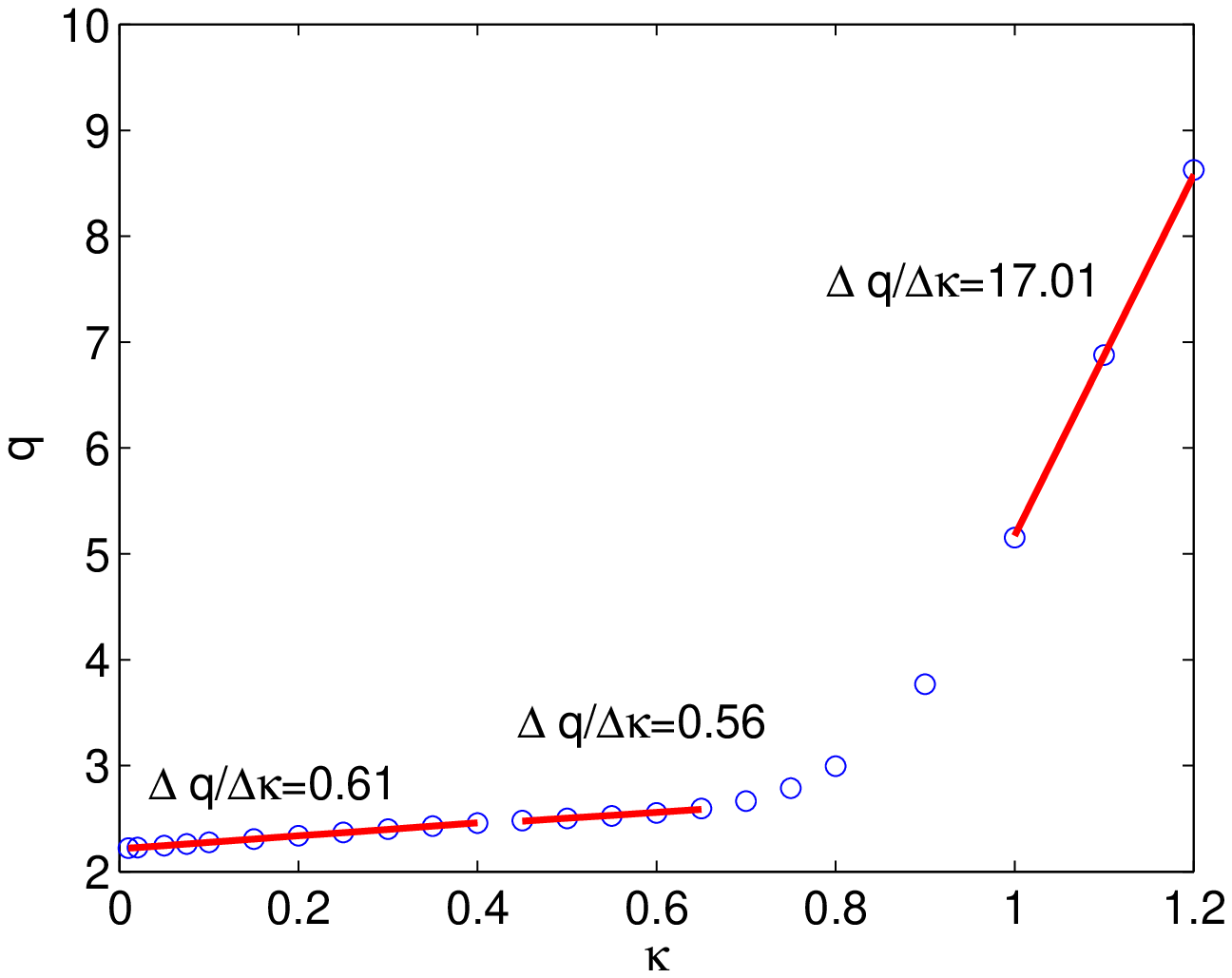}}}}
 \caption{Possible linear relationship between the power-law exponent $q$ and $\kappa$}
\label{Fig5}
\end{figure}

\newpage

\begin{figure} [htb]
\centerline{{\scalebox{0.55} {\includegraphics{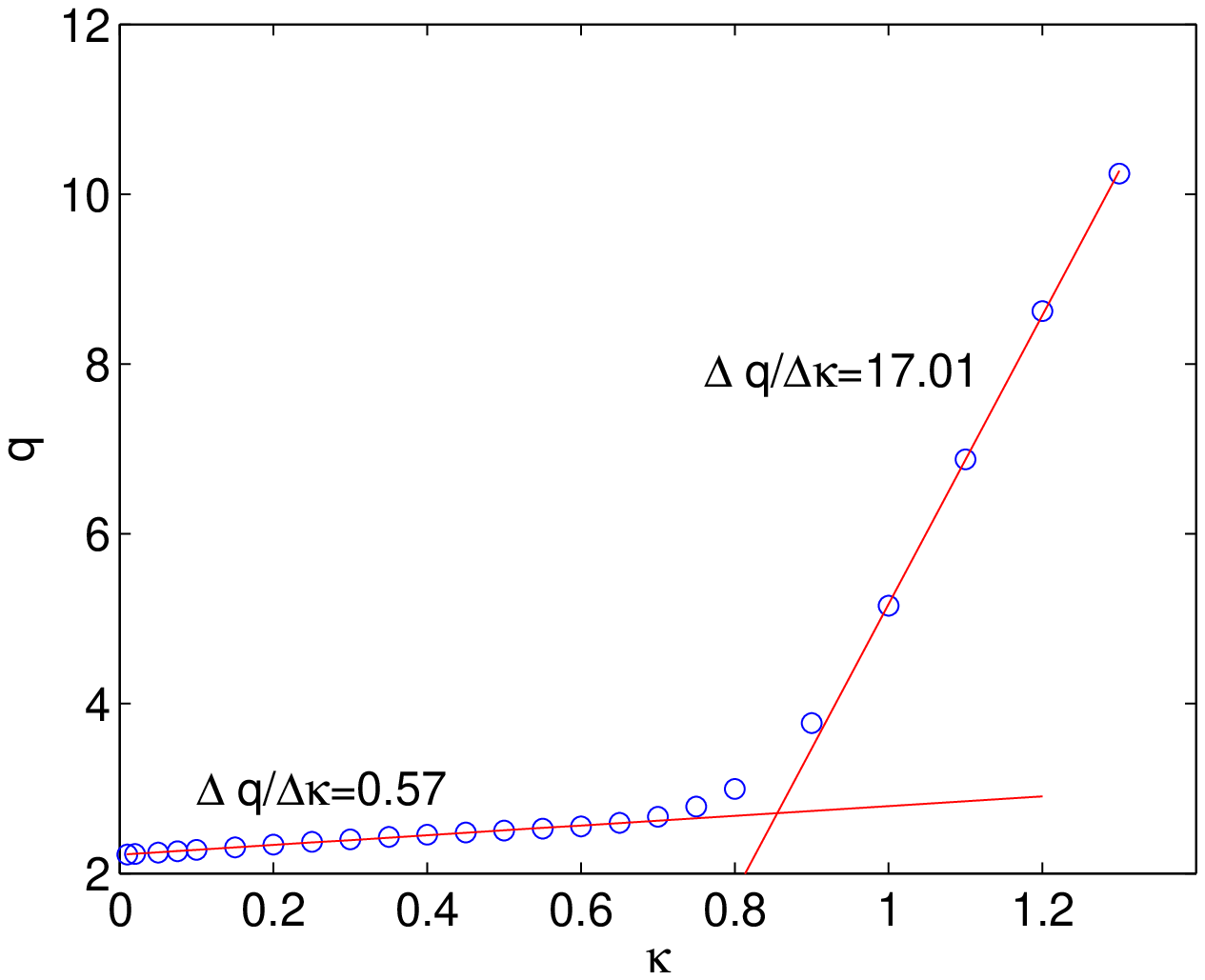}}}}
 \caption{Possible linear
relationship between the power-law exponent $q$ and $\kappa$}
\label{Fig6}
\end{figure}

\newpage

\begin{figure} [htb]
\centerline{{\scalebox{0.55} {\includegraphics{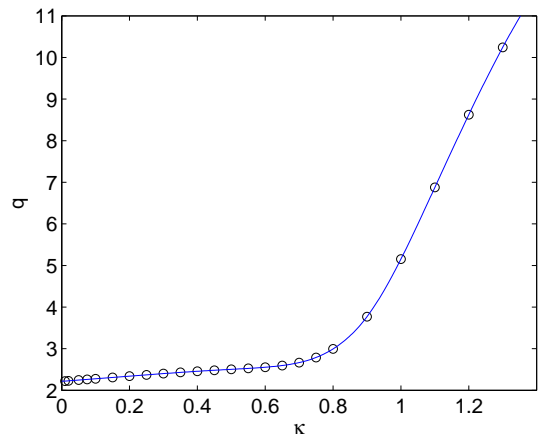}}}}
 \caption{Spline relationship between the power-law exponent $q$ and $\kappa$}
\label{Fig7}
\end{figure}

\newpage

\begin{figure} [htb]
\centerline{{\scalebox{0.55} {\includegraphics{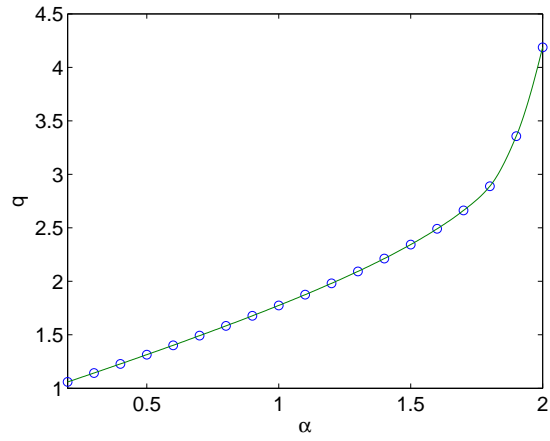}}}}
 \caption{Fractional derivative order $\alpha$ vs. power law exponent $q$
  adjusting the solutions of the fractional advection-diffusion equation ( $\kappa=0.2$)}
\label{Fig8}
\end{figure}

\newpage

\begin{figure} [htb]
\centerline{{\scalebox{0.55} {\includegraphics{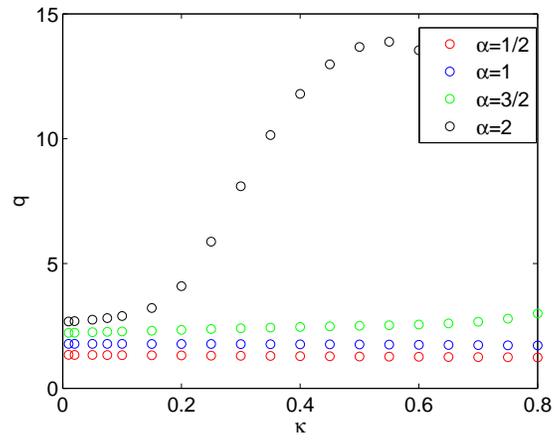}}}}
 \caption{$q$ vs. $\kappa$ for different $\alpha-$values.}
\label{Fig9}
\end{figure}

\end{document}